\documentclass[a4paper]{article}

\usepackage[pdftex]{graphicx}
\usepackage{graphicx}
\graphicspath{{figures/}}
\usepackage{color}
\usepackage{fancyhdr}
\usepackage{authblk}
\usepackage{booktabs}
\usepackage{tabularx}
\usepackage{float}
\usepackage{placeins}
\usepackage[dvipsnames]{xcolor}
\usepackage[
    colorlinks=true,
    linkcolor=NavyBlue,
    citecolor=NavyBlue,
    urlcolor=NavyBlue
]{hyperref}

\title{Deep Learning for Accelerated Long-Horizon Forecasting of Multicomponent Multiphase Microstructure Evolution in High-Entropy Alloys}

\author{
Hamidreza Razavi\\
hamidreza.razavi@kuleuven.be
\and
Nele Moelans\\
nele.moelans@kuleuven.be}

\affil{Department of Materials Engineering, KU Leuven\\
Kasteelpark Arenberg 44 Bus 2450, 3001 Leuven, Belgium}

\date{}

\newcommand{\ud}{\mathrm{d}}

\begin{document}

\maketitle

\begin{abstract}
Phase-field modeling provides a powerful approach for predicting microstructure evolution but becomes computationally prohibitive for multicomponent and multiphase systems over large spatial and temporal scales. This work presents an autoencoder--graph convolutional network--long short-term memory (AE--GCN--LSTM) surrogate framework for long-horizon forecasting of microstructure evolution in the multicomponent AlCrFeNi high-entropy alloy system containing coexisting BCC and FCC phases. A multi-head autoencoder compresses the four elemental concentration fields and phase-field order parameter into latent representations, which are formulated as graphs for learning their spatial and temporal evolution. The framework accurately forecasts microstructure evolution over horizons extending to 3,000,000 simulation timesteps. Its robustness is systematically evaluated under previously unseen conditions without retraining, fine-tuning, or parameter adaptation. These evaluations include variations in FCC precipitate size and initial position, microstructures containing one, two, and five FCC precipitates, and complex phase interactions involving precipitate merging and splitting. Although trained only on $100 \times 100$ computational domains containing a single nominal alloy composition, the framework is successfully transferred to larger $256 \times 256$ and $512 \times 512$ systems and to previously unseen AlCrFeNi compositions. Across the evaluated configurations, the model preserves the dominant phase morphology and compositional evolution while providing computational speedups ranging from approximately $7.2 \times 10^{3}$ to $6.23 \times 10^{4}$ relative to conventional phase-field simulations. These results demonstrate that latent graph-based AE--GCN--LSTM forecasting provides a scalable and computationally efficient surrogate for long-horizon simulation of multicomponent, multiphase microstructures and offers a promising foundation for high-throughput alloy design.
\end{abstract}


\section{Introduction}\label{sec1}

The evolution of microstructures plays a significant role in the computational design of materials as it governs the mechanical, thermal, and chemical performance of engineering alloys. Phase-field modeling remains one of the most powerful methods for predicting the evolution of complex microstructures. This technique naturally captures diffusion, phase transformations, precipitate nucleation, growth, coarsening, merging, and splitting without explicitly tracking moving interfaces \cite{Moelans2008}. Despite its predictive capacity, conventional phase-field simulation has its own drawbacks, such as high computational cost, particularly for multicomponent and multiphase \cite{Zakaria}. This is especially the case where multiple coupled concentration fields and order parameters must be solved over extended temporal and spatial scales. These challenges pose a significant obstacle to high-throughput alloy design and optimization. Therefore, it is paramount to develop efficient surrogate models that accurately reproduce phase-field evolution at a fraction of the computational cost and time.

Recent advances in artificial intelligence have given rise to numerous new techniques that hold promise for accelerating phase-field simulations through data-driven approaches. They have demonstrated considerable promise for accelerating phase-field simulations through data-driven surrogate modeling. Gong et al. developed a machine learning surrogate framework for multicomponent alloy solidification. However, it focuses on solidification rather than direct long-horizon forecasting of microstructure evolution \cite{Hao}. Zong et al. had developed a deep-learning surrogate to learn microstructure evolution in titanium alloys from phase-field data. The work does not address multicomponent HEAs phase separation and also does not address generalization across compositions or spatial resolutions \cite{Zong}. Ahmad et al. implement an autoencoder plus attention ConvLSTM for the prediction of ternary spinodal dealloying and late-stage coarsening. Their work  is limited to ternary alloys and is based on hybrid phase-field and deep learning prediction. Therefore, it lacks direct long-horizon forecasting in multicomponent high-entropy alloys \cite{Ahmad}. Subedi et al. propose a ConvLSTM framework which predicts 2D microstructure frames in image space and uses phase-field data. Despite the remarkable predictions accuracy, the framework is not focused on multicomponent HEA \cite{Subedi}. GrainGNN works on quantifying graph-based grain and is performed in full-resolution microstructures. The graphs are built from pixel grids and do not implement dimensionality reduction, which is helpful for a shorter training time \cite{graingnn}. Tiwari et al. work accelerate phase-field simulation by employing convolutional recurrent networks. In their work, training is performed directly on full-resolution fields, which limits scalability to multicomponent HEAs \cite{tiwari}. Peivaste et al. uses U-Net architectures for the prediction of microstructure evolution. Although the approach is promising, it is limited to single-step predictions and is not applied to HEAs \cite{iman} . 

Although all of these works contribute greatly to the advancement of machine learning based surrogate modeling, the majority of existing approaches have focused on binary alloy systems, short-term forecasting, or predictions confined to the training data distribution. The extension of these surrogate models to realistic multicomponent high-entropy alloys (HEAs), multiphase microstructures, and long-horizon evolution remains largely unexplored. Furthermore, the ability of these models to generalize beyond the training distribution, such as unseen precipitate morphologies, spatial configurations, and computational domain sizes, has received limited attention.

Previously, we developed a latent-space graph neural network framework for long-horizon forecasting of binary phase-field microstructure evolution \cite{Razavi-Moelans}. A latent physics-informed GCN–LSTM framework was developed for long-horizon forecasting of microstructure evolution in the Bismuth–Antimony (Bi–Sb) alloy. The model successfully generalized to unseen microstructures without retraining and accurately forecasted evolutions up to 40,000 timesteps. In this work, we present an extended surrogate framework to accelerate phase-field simulations of the multicomponent AlCrFeNi high-entropy alloy that contains coexisting BCC and FCC phases. The proposed framework combines an autoencoder with a graph convolutional network and long short-term memory (GCN--LSTM) architecture to learn the temporal evolution of compressed latent microstructure representations. Once trained, it enables accurate long-horizon forecasting of the evolution of multicomponent microstructures up to 3,000,000 timesteps. The framework provides substantial computational acceleration compared to conventional phase-field simulations with a high predictive accuracy over extended temporal horizons. The results obtained make this proposed framework an attractive option for significantly accelerating multicomponent high entropy alloy design. 

Extending the binary framework to realistic multicomponent HEAs introduces its own set of challenges and complexities. Such complexities arise due to the coupled evolution of multiple elemental concentration fields, multiple coexisting phases, and increasingly complex morphological interactions. In particular, the simultaneous evolution of BCC and FCC phases together with four elemental concentration fields presents a significantly more challenging forecasting problem than binary phase separation. A central objective of this work is to assess whether the learned evolution dynamics remain transferable under conditions that differ substantially from those used for training. Therefore, the model is subjected to systematic zero-shot generalization studies spanning increasingly challenging spatial, morphological, and compositional variations. These evaluations are specifically designed to determine whether the surrogate captures the underlying phase-evolution physics rather than memorizing training trajectories. This allows a more rigorous assessment of model robustness and practical applicability to realistic multicomponent alloy design problems. The proposed framework exhibits strong zero-shot generalization capabilities to previously unseen precipitate sizes, spatial locations, precipitate counts, and increasingly complex FCC phase morphologies. The generalization capabilities also include precipitate merging and splitting, without retraining or parameter adaptation. Furthermore, although training was exclusively performed on $128 \times 128$ phase-field simulations, the model successfully transfers to previously unseen computational domains of $256 \times 256$ and $512 \times 512$. Finally, despite being trained using microstructures generated from a single nominal AlCrFeNi alloy composition, the model accurately forecasts the evolution of multiple previously unseen alloy compositions without retraining. These characteristics demonstrate the scalability and robustness of the learned evolution dynamics across different simulation parameter settings, showing that the learned dynamics remain transferable to variations in alloy chemistry.

\section{Results}\label{sec2}

\subsection{Long-Horizon Forecasting on Held-Out Validation Set}

Following training, the model was evaluated on a held-out validation set of previously unseen microstructures. The forecasting model was provided with multiple previous microstructure states as input, where each state was represented by five channels consisting of the four elemental concentration fields and the phase-field variable. The input sequence was then encoded using the trained multi-head autoencoder to obtain a sequence of latent representations. These latent states were converted into graph-structured inputs and provided to the trained GCN--LSTM forecasting model. The forecasting model then predicted the future latent evolution of the elemental concentration fields and phase-field variable while preserving the conservation constraints learned during training.

The predicted elemental concentration fields and the phase-field variable were reconstructed in the physical domain using the decoder. The reconstructed predictions were compared with the corresponding unseen phase-field simulation data. The prediction accuracy was evaluated using the mean squared error (MSE) and structural similarity index (SSIM) for the elemental concentration fields, and binary cross-entropy (BCE) to assess phase-field prediction accuracy. This workflow enables long-horizon prediction of future microstructure states from a limited sequence of prior observations while evaluating the model's ability to generalize across unseen test microstructures.

The model's best performing epoch is 94 with a mean physical-space MSE of $8.057e-06$, a mean SSIM of $0.9993$, a phase-field BCE of $2.460e-10$, and a conservation loss of $4.497e-16$ on the held-out validation set. 

\begin{figure}[H]
    \centering
    \includegraphics[width=1.0\linewidth]{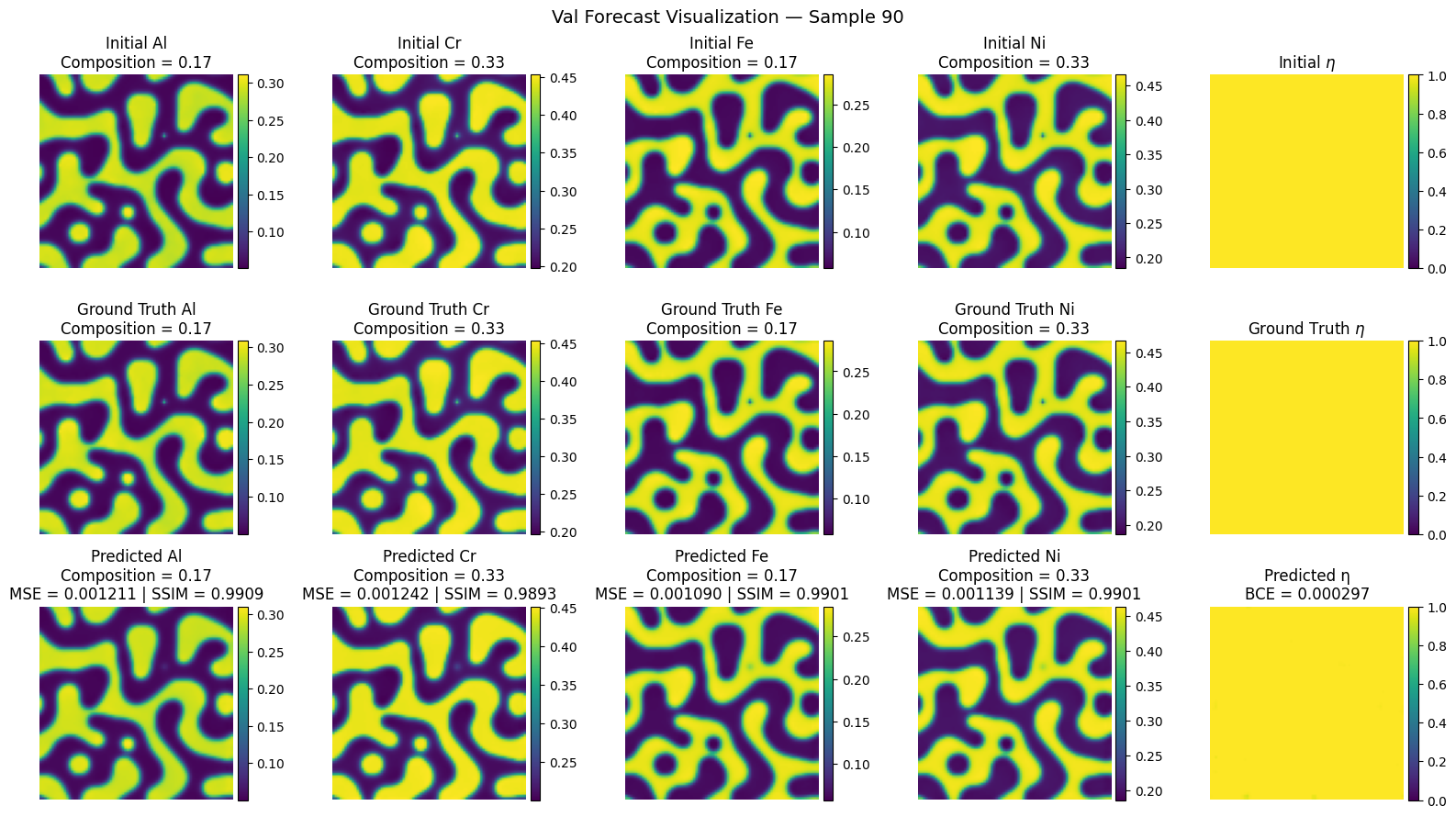}
    \caption{Validation forecasting metrics during model training}
    \label{fig:model_training}
\end{figure}

\subsection{Generalization to Unseen Microstructure Configurations}

Additional forecasting experiments were performed on phase-field simulations generated from microstructure configurations absent from the training, validation, and test datasets. The training dataset consisted primarily of microstructures containing a single FCC precipitate morphology. The unseen datasets were introduced gradually with larger deviations from the training conditions. The variations consisted of precipitate size, location, count, spatial arrangement, and overall morphological complexity.

For each unseen configuration, a short sequence of early microstructure states (3 frames) was provided as a forecasting context. Meanwhile, all other model parameters remained fixed without any retraining, fine-tuning, transfer learning, or parameter adaptation. The trained autoencoder and GCN--LSTM framework were applied directly to the new microstructures, and future states were predicted using the same procedure employed during standard forecasting.

The generalization study included configurations that contained one single small FCC precipitates, one single large FCC precipitates, dual FCC precipitates, and five FCC precipitates. Collectively, these configurations provide a systematic assessment of the model's ability to generalize to previously unseen morphologies and increasingly complex precipitate interactions while maintaining accurate long-horizon forecasting performance.

\subsubsection{Single Small FCC Precipitate Configuration}

The first generalization study considered a microstructure containing a single small FCC precipitate positioned near the lower region of the computational domain. This configuration differs from the training data in both precipitate size and spatial placement. Forecasts were generated only from the initial evolution history, and predictions were made without retraining with the frozen model.

\begin{figure}[H]
    \centering
    \includegraphics[width=\linewidth]{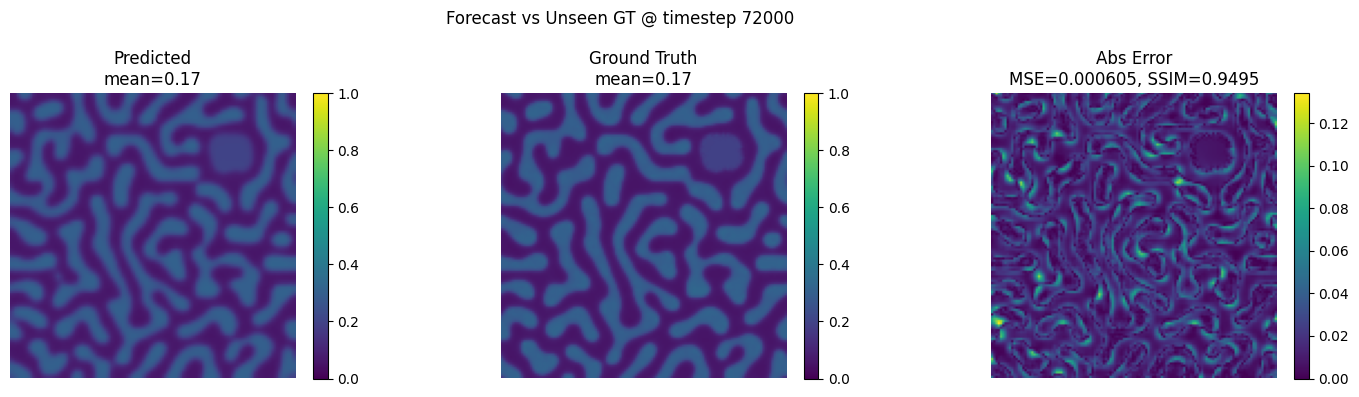}
    \includegraphics[width=\linewidth]{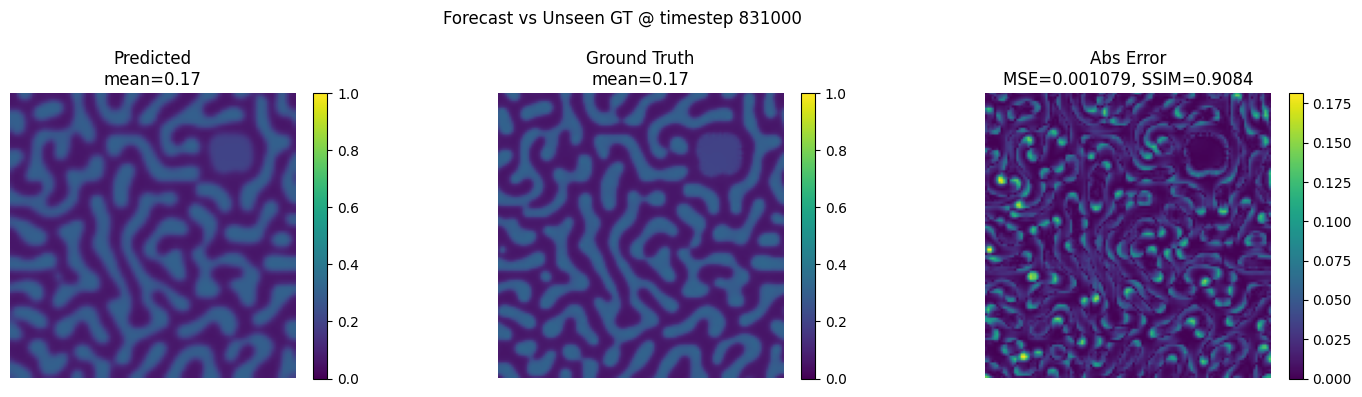}
    \includegraphics[width=\linewidth]{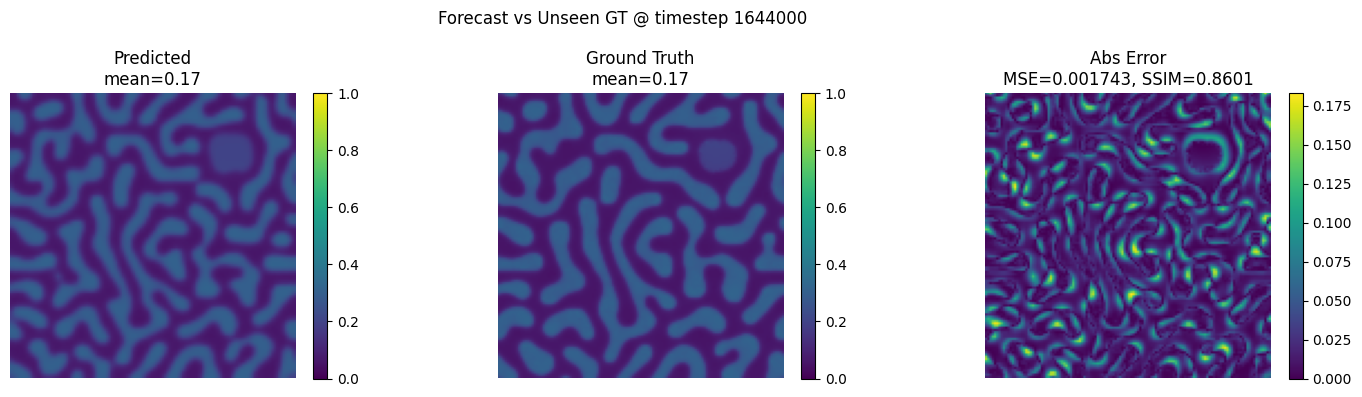}
    \caption{Single small FCC forecasting up to $1,\!644,\!000$ timesteps. }
    \label{fig:small_fcc}
\end{figure}

\subsubsection{Single Large FCC Precipitate Configuration}

The second study examined a microstructure containing a significantly larger FCC precipitate Relative to the training configurations. This dataset introduces a considerably different phase morphology and precipitate volume fraction. This configuration evaluates whether the latent-space forecasting framework can accurately capture long-term evolution when the characteristic microstructural length scales differ from those encountered during training.

\begin{figure}[H]
    \centering
    \includegraphics[width=\linewidth]{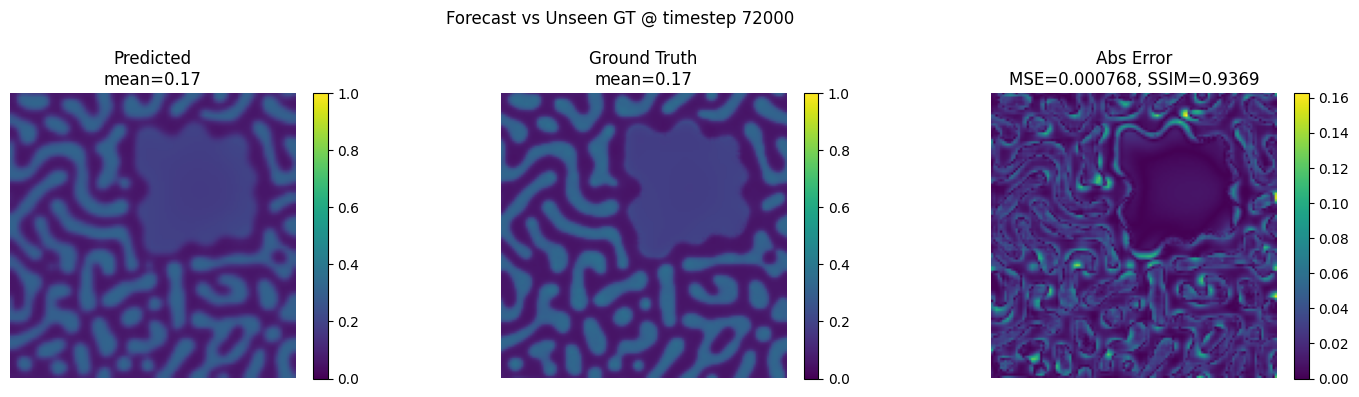}
    \includegraphics[width=\linewidth]{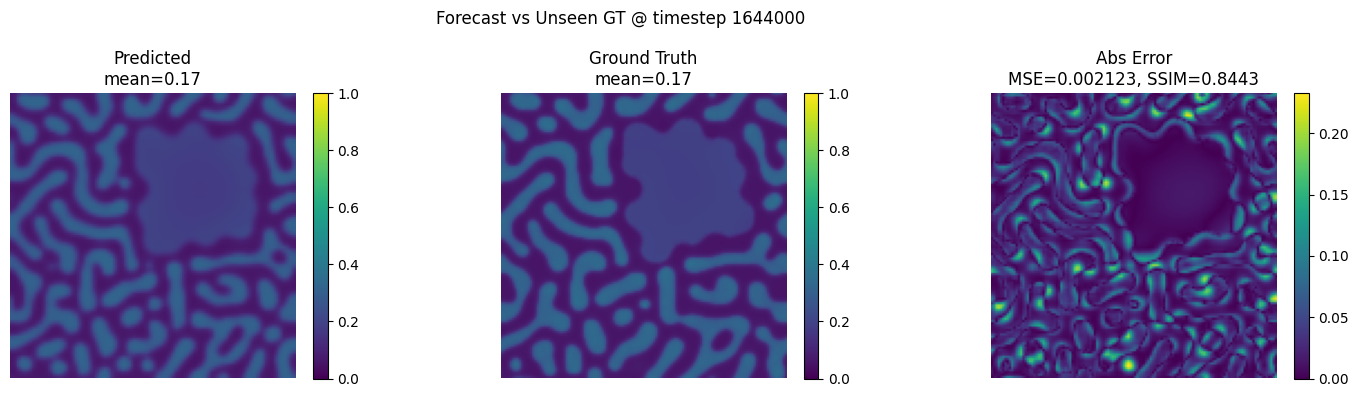}
    \includegraphics[width=\linewidth]{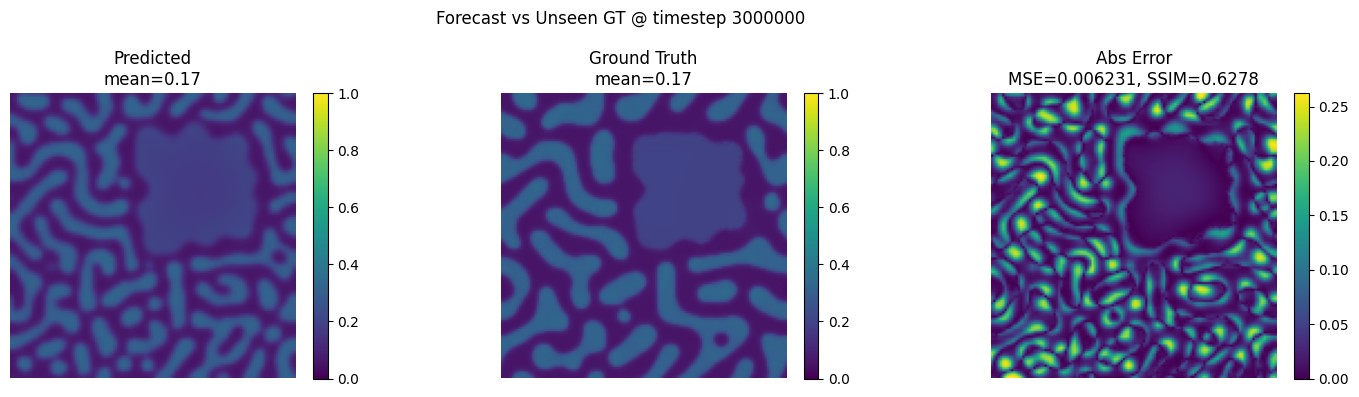}
    \caption{Single large FCC forecasting up to $3,\!000,\!000$ timesteps. }
    \label{fig:large_fcc}
\end{figure}

\subsubsection{Dual FCC Precipitate Configuration}

The third generalization study considered a microstructure containing two FCC precipitates. This configuration introduces additional phase interfaces and more complex spatial interactions compared to the single-precipitate cases. Since the model was not exposed to this morphology during training, this dataset provides a stringent assessment of the framework's ability to generalize to more complex precipitate distributions.

\begin{figure}[H]
    \centering
    \includegraphics[width=\linewidth]{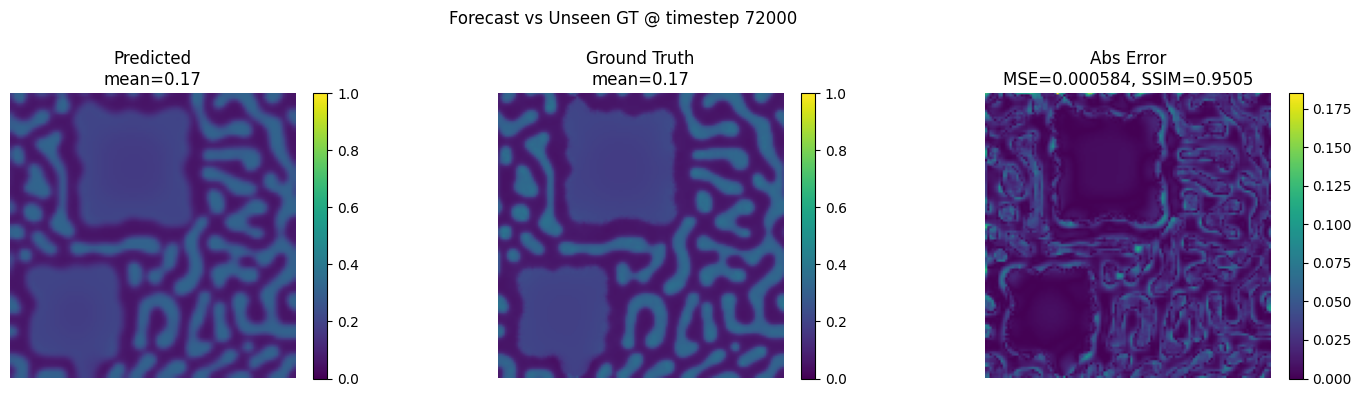}
    \includegraphics[width=\linewidth]{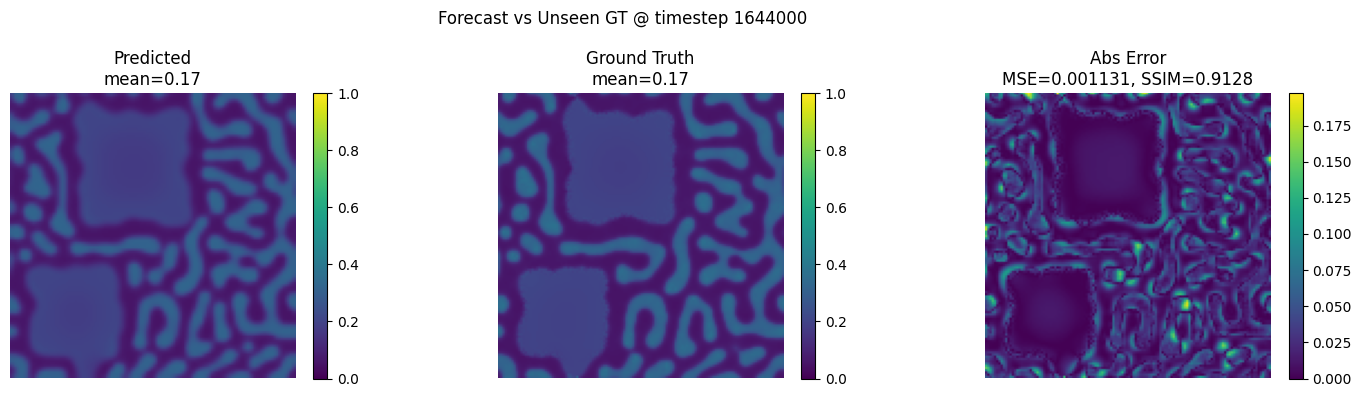}
    \includegraphics[width=\linewidth]{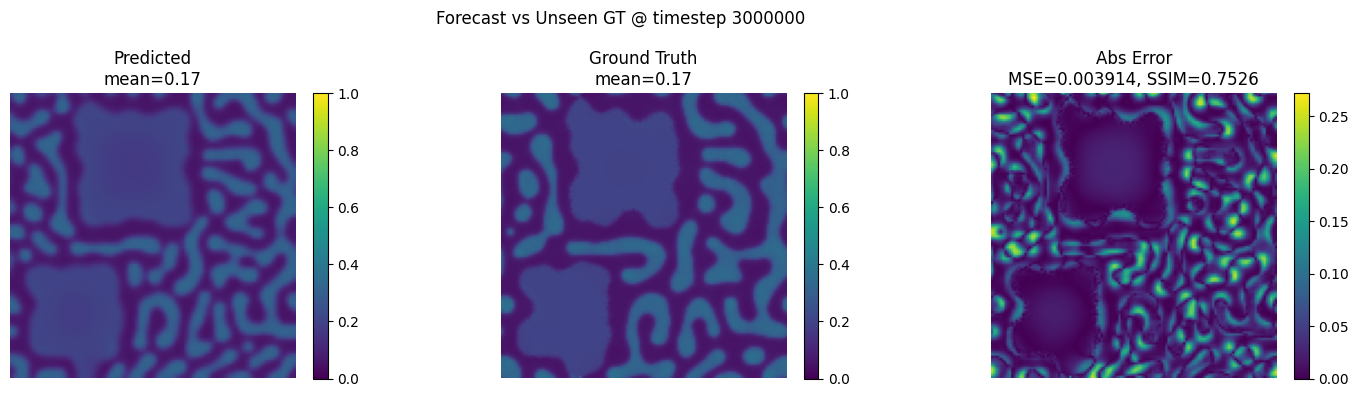}
    \caption{Dual FCC FCC forecasting up to $3,\!000,\!000$ timesteps. }
    \label{fig:two_fcc}
\end{figure}

\subsubsection{Five FCC Precipitate Configuration}

The final study considered an unseen microstructure containing five FCC precipitates distributed throughout the computational domain. Compared with previous configurations, this dataset exhibits substantially greater morphological complexity due to an increased number of precipitates and phase interfaces.

This configuration evaluates whether the latent-space representation and learned temporal dynamics remain effective when the number of precipitates and the overall microstructural complexity increase significantly. Forecasting performance was assessed over multiple future time steps by comparing the predicted concentration and phase distributions with the corresponding ground-truth phase-field simulations.

\begin{figure}[H]
    \centering
    \includegraphics[width=\linewidth]{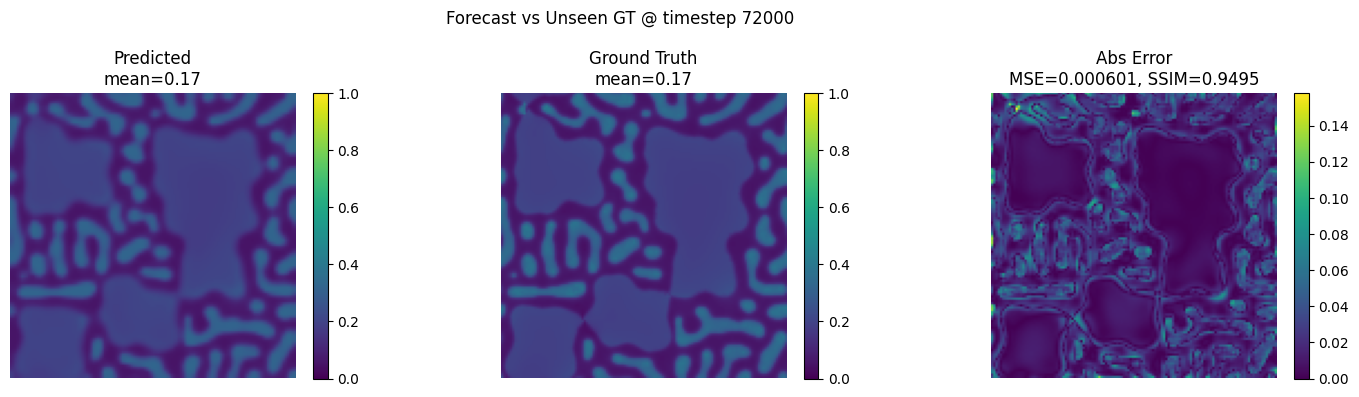}
    \includegraphics[width=\linewidth]{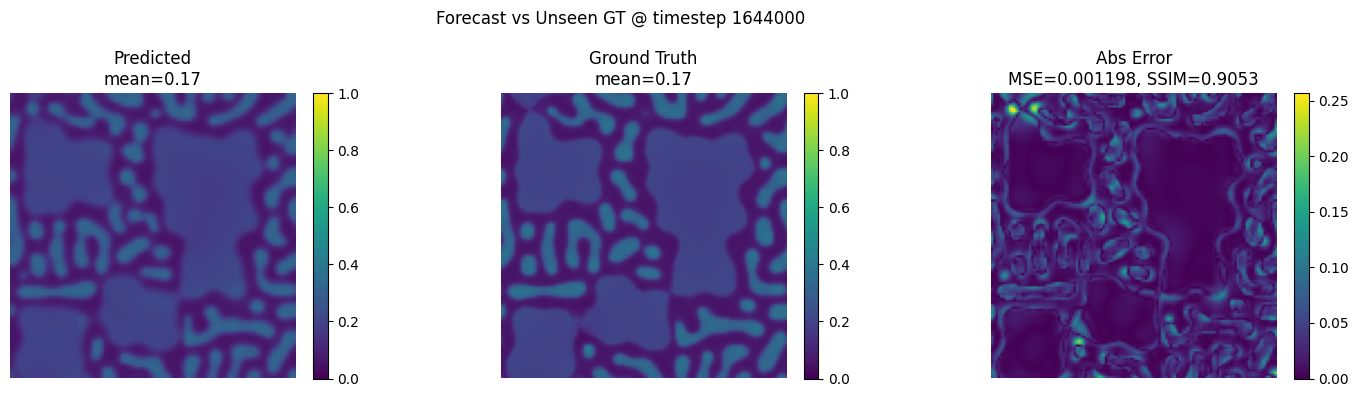}
    \includegraphics[width=\linewidth]{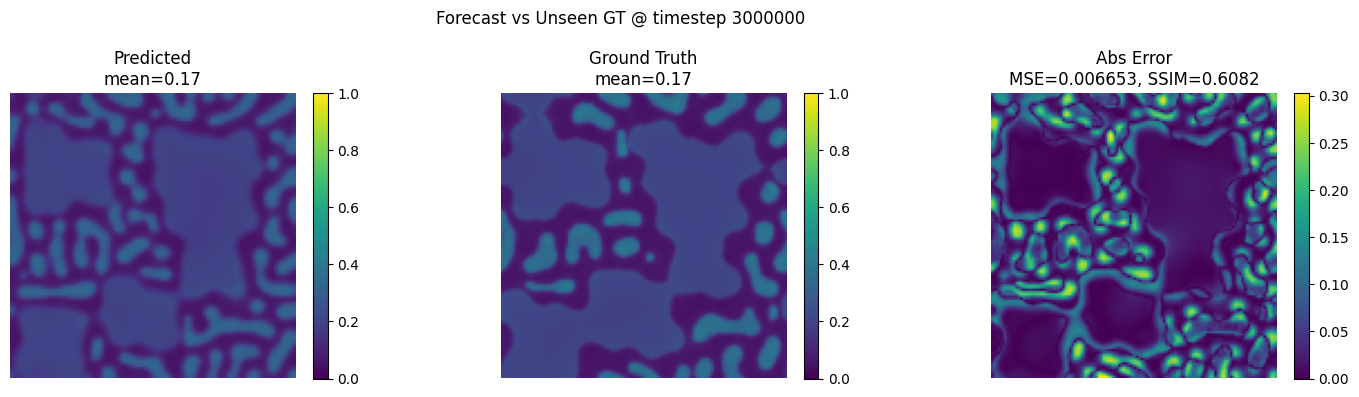}
    \caption{Five FCC forecasting up to $3,\!000,\!000$ timesteps. }
    \label{fig:five_fcc}
\end{figure}

\subsection{Generalization to Unseen Spatial Resolutions}

In addition to morphological variations, the scalability of the proposed framework was evaluated on microstructures with spatial resolutions that were not encountered during training. Although model development was performed using microstructures generated on a fixed spatial domain with dimensions $128 \times 128$, practical materials design applications often require forecasting on larger computational domains.

To investigate this capability, forecasting experiments were performed on unseen $256 \times 256$ and $512 \times 512$ microstructures without retraining. Because direct application of the trained model to larger domains resulted in reduced spatial fidelity, a patch-based forecasting strategy was employed. The patch-based strategy enables the learned latent-space evolution dynamics to be transferred to larger computational domains while preserving the original model parameters. The resulting experiments provide a rigorous assessment of the framework's ability to generalize across spatial scales and maintain forecasting accuracy beyond the domain sizes encountered during training.

\subsubsection{Generalization to a \texorpdfstring{$256 \times 256$}{256 x 256} Domain}

To further evaluate the scalability of the framework, forecasting experiments were performed on an unseen $256 \times 256$ microstructure domain. Since the model was initially trained on $128 \times 128$ microstructures, initial attempts on unseen $256 \times 256$ failed due to blurry predictions. 

To address this limitation without retraining, the $256 \times 256$ microstructure was decomposed into overlapping $128 \times 128$ subdomains. Each subdomain was processed independently using the frozen autoencoder and the GCN--LSTM framework. Subsequently, the resulting patch-level forecasts were combined by weighted blending to preserve spatial continuity and minimize stitching artifacts. This strategy enables the learned dynamics of microstructure evolution to be transferred to larger spatial domains while maintaining the original model parameters. 

\begin{figure}[H]
    \centering
    \includegraphics[width=\linewidth]{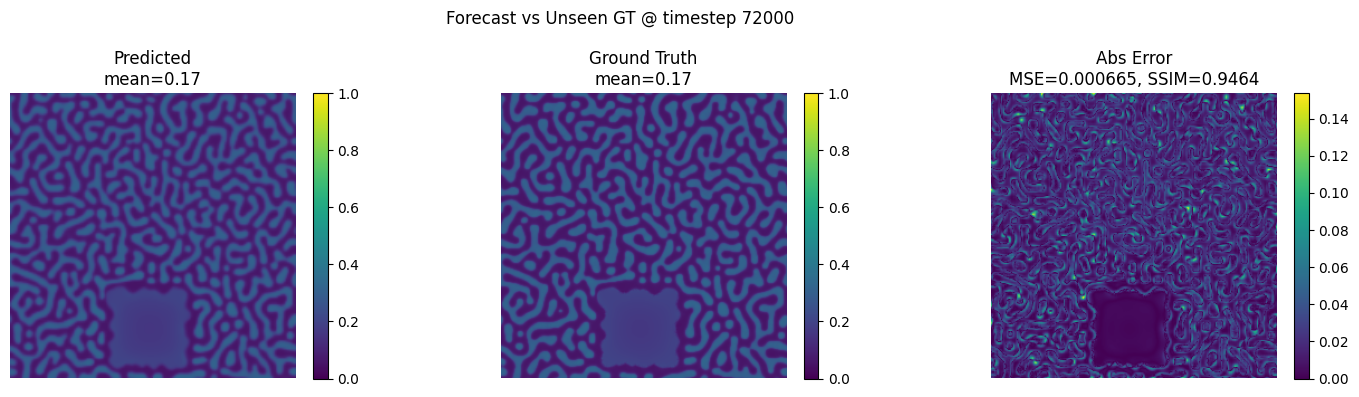}
    \includegraphics[width=\linewidth]{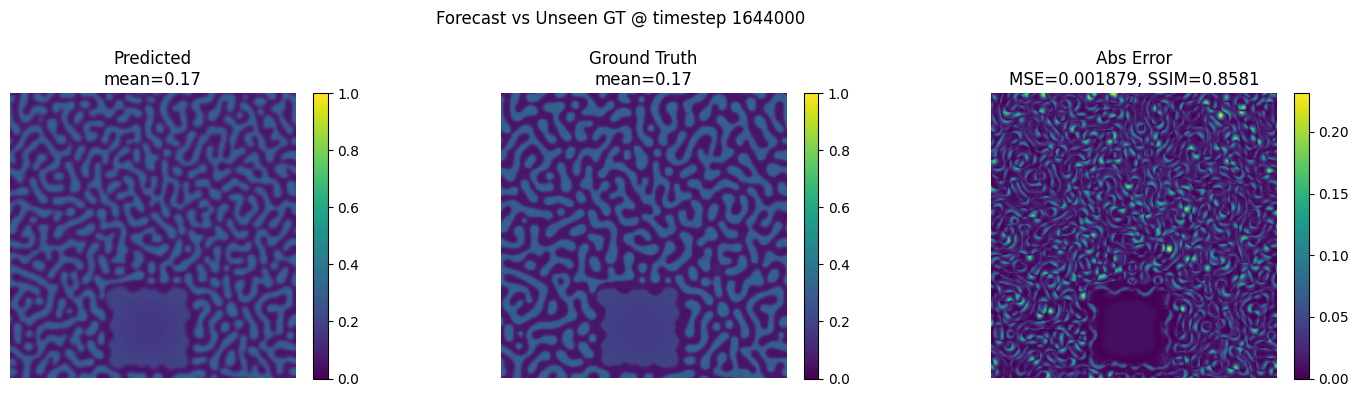}
    \includegraphics[width=\linewidth]{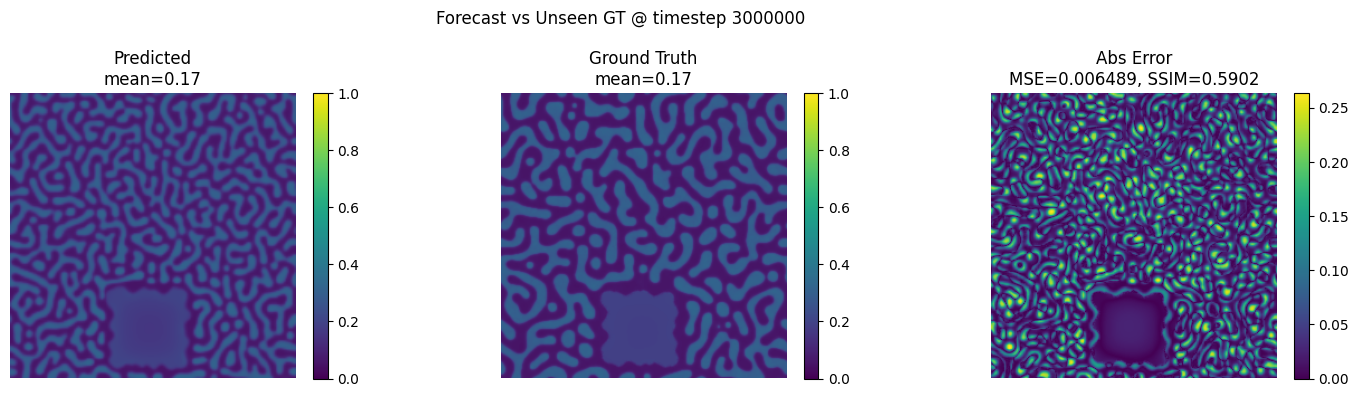}
    \caption{Forecasting of $256 \times 256$ up to $3,\!000,\!000$ timesteps. }
    \label{fig:resolution_a}
\end{figure}

\subsubsection{Generalization to a \texorpdfstring{$512 \times 512$}{512 x 512} Domain}

The same patch-based forecasting strategy from the previous study was further applied to an unseen $512 \times 512$ microstructure. This facilitates evaluating whether the learned latent evolution dynamics remains effective at substantially larger spatial scales. The forecasting was performed using the same frozen autoencoder and GCN--LSTM framework, without retraining or parameter modifications.

For this study, the resulting prediction was compared with multiple future ground-truth states, as the $512 \times 512$ simulation exhibits greater morphological complexity and evolves over a substantially longer temporal window. To identify the most closely matching microstructure state, a best-match scan was performed.

Successful forecasting in both the $256 \times 256$ and $512 \times 512$ domains demonstrates that the latent dynamics are not restricted to the training spatial dimensions. Therefore, the framework can be transferred to larger microstructures without retraining, which highlights its scalability and robustness.

\begin{figure}[H]
    \centering
    \includegraphics[width=\linewidth]{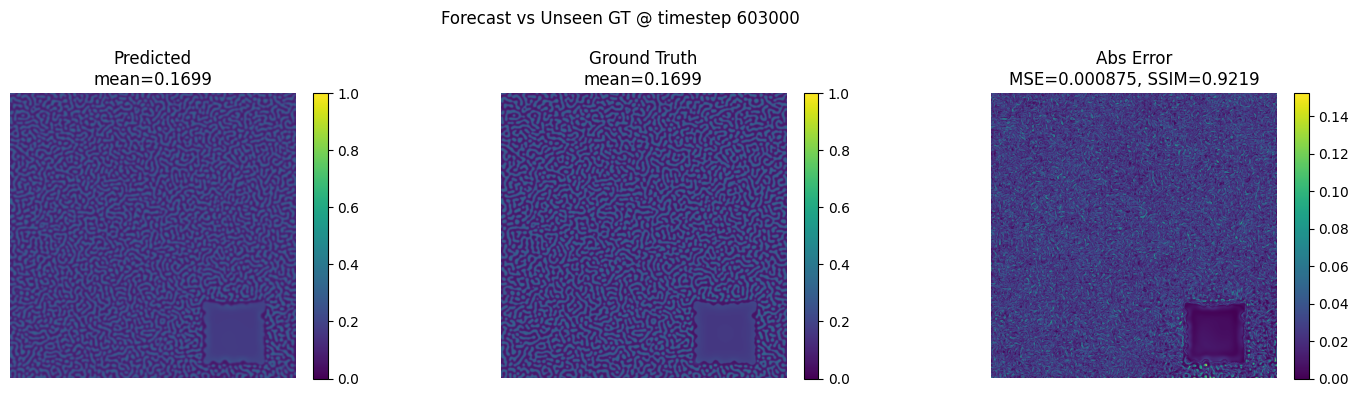}
    \includegraphics[width=\linewidth]{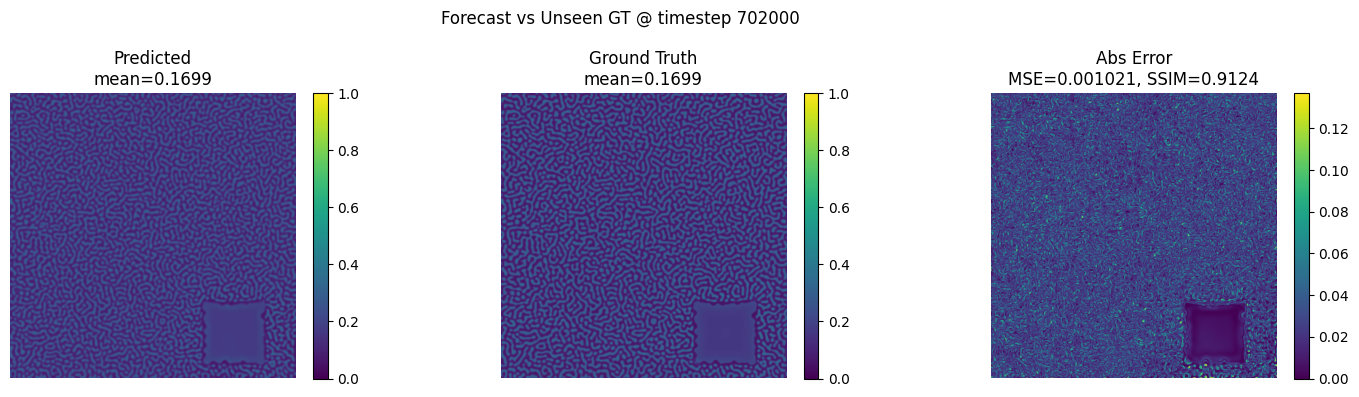}
    \includegraphics[width=\linewidth]{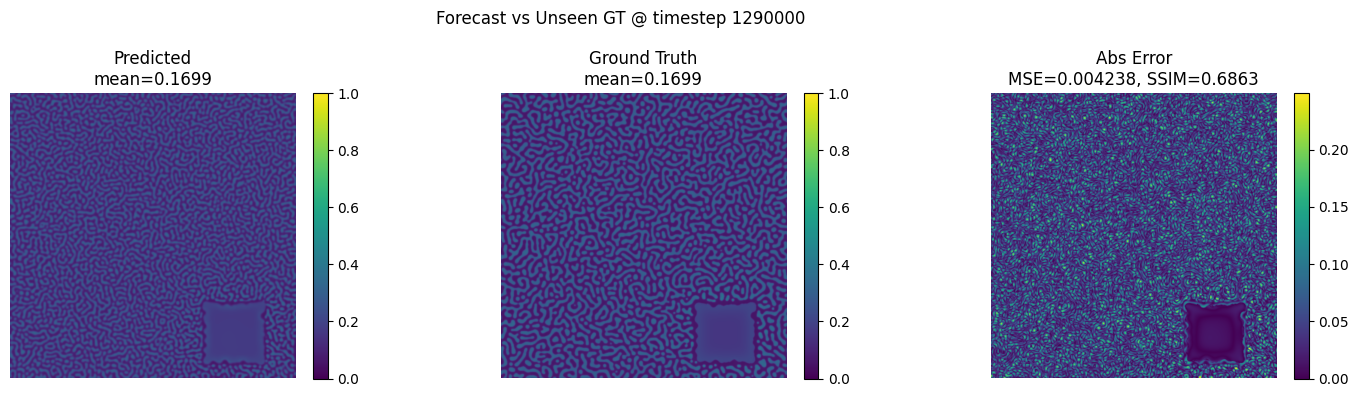}
    \caption{Forecasting of $512 \times 512$ up to $1,\!290,\!000$ timesteps. }
    \label{fig:resolution_b}
\end{figure}

\subsection{Generalization to Unseen Alloy Compositions}

This section assesses whether a model trained on a single nominal composition can accurately predict the evolution of the microstructure under an unseen alloy composition. The forecasting model was trained exclusively using microstructures generated at the nominal alloy composition of Al$=0.17$, Cr$=0.33$, Fe$=0.17$, and Ni$=0.33$. To evaluate the transferability of the learned latent evolution dynamics across variations in alloy chemistry, forecasting experiments were performed on two previously unseen compositions. The first corresponded to Al$=0.18$, Cr$=0.32$, Fe$=0.18$, and Ni$=0.32$ ($+0.01$ Al/Fe, $-0.01$ Cr/Ni), followed by a larger variation of Al$=0.19$, Cr$=0.31$, Fe$=0.19$, and Ni$=0.31$ ($+0.02$ Al/Fe, $-0.02$ Cr/Ni).

\subsubsection{\texorpdfstring{$\pm 0.01$}{+/- 0.01} Composition Variation}

In this experiment, predictions were generated using the frozen autoencoder and GCN–LSTM model without retraining by $\pm0.01$ variation. Based on the results, the predictions preserved the correct global morphology, accurately captured the location of the FCC precipitate, and maintained the overall compositional balance. However, compared with the corresponding ground-truth phase-field simulations, the predicted spinodal structures exhibited smoother phase boundaries and reduced high-frequency spatial detail. 

\begin{figure}[H]
    \centering
    \includegraphics[width=\linewidth]{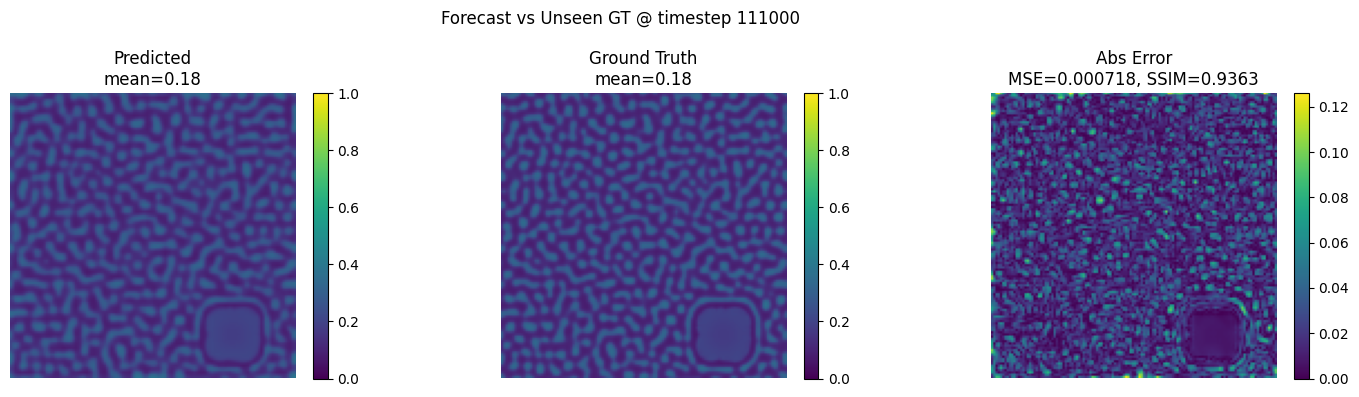}
    \includegraphics[width=\linewidth]{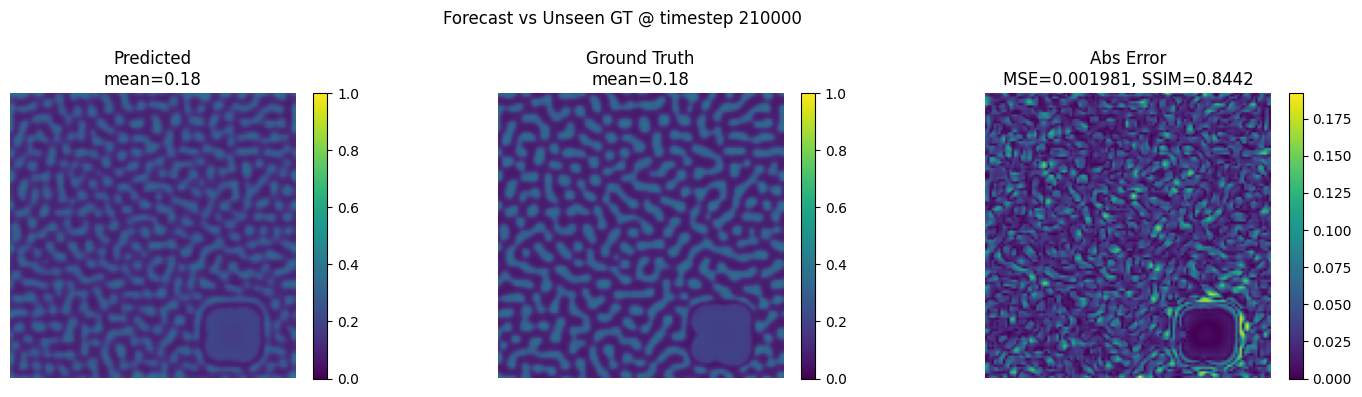}
    \includegraphics[width=\linewidth]{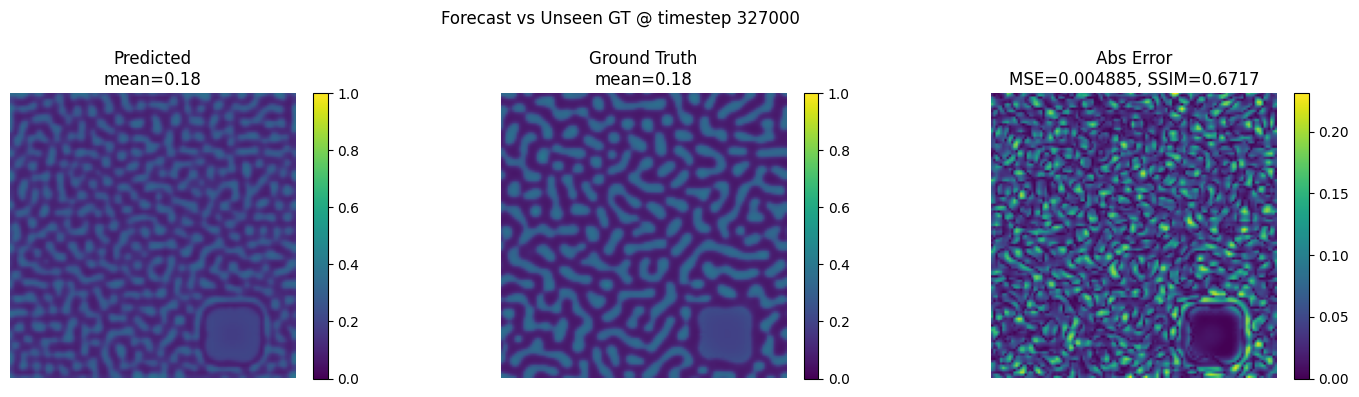}
    \caption{Forecasting under unseen $\pm1\%$ composition variation. }
    \label{fig:composition_a}
\end{figure}

\subsubsection{\texorpdfstring{$\pm 0.02$}{+/- 0.02} Composition Variation}

For the $\pm0.02$ variations, the resulting forecasts preserved the dominant precipitate morphology and the overall compositional balance. However, relative to the $\pm0.01$ perturbation, the predicted microstructure exhibited a greater reduction in fine-scale spatial fidelity. This deviation presents increasingly smoothed spinodal features compared with the corresponding phase-field simulations. These observations suggest that although the proposed framework remains capable of reproducing the overall evolution under unseen alloy compositions, progressively larger deviations from the training chemistry reduce local prediction accuracy.

\begin{figure}[H]
    \centering
    \includegraphics[width=\linewidth]{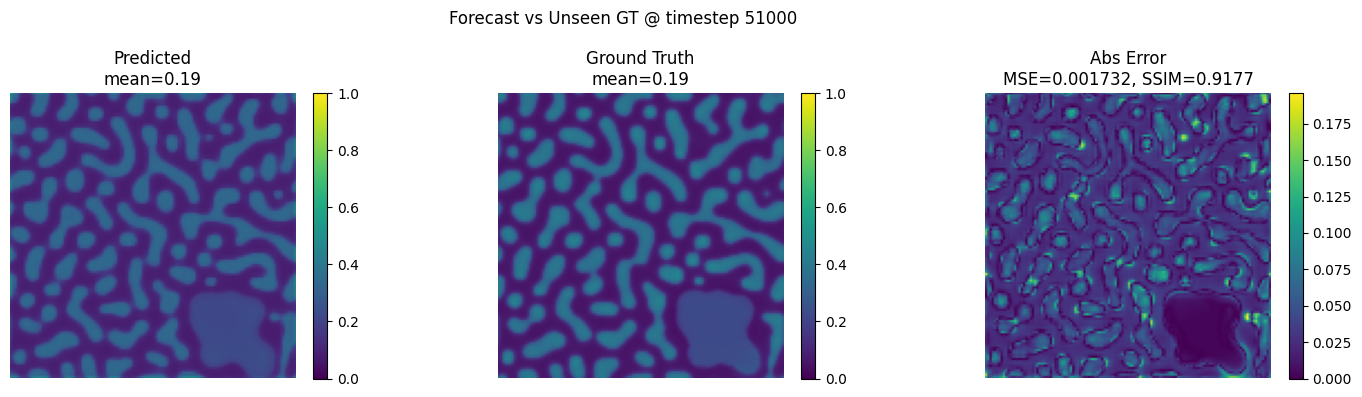}
    \includegraphics[width=\linewidth]{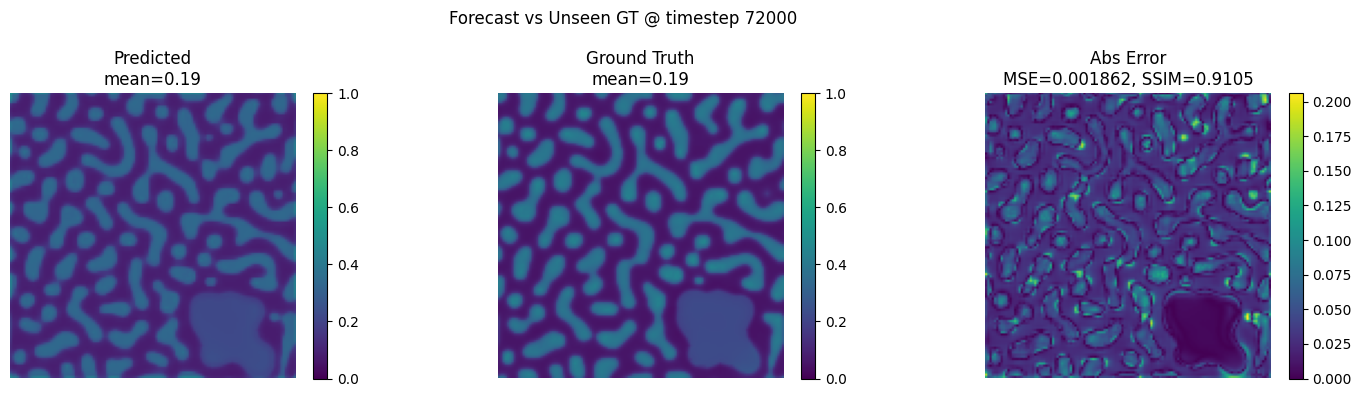}
    \includegraphics[width=\linewidth]{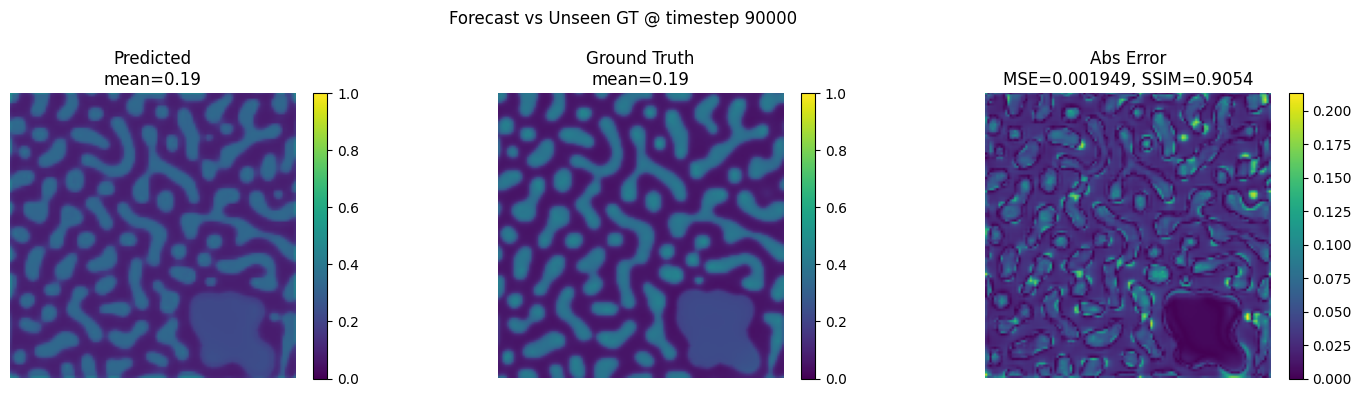}
    \caption{Forecasting under unseen $\pm2\%$ composition variation.}
    \label{fig:composition_b}
\end{figure}

\subsection{Runtime Comparison}

Table~\ref{tab:runtime_comparison} Compares the computational times  it takes to perform phase-field simulations versus the corresponding time of GCN--LSTM forecasts. Most of the simulations were performed up to timestep 3,000,000. However, since the $512 \times 512$ simulation requires substantially greater computational time, both the forecasting and simulation were evaluated up to timestep 1,290,000. On the other hand, the evaluation of $\pm 0.01$ and $\pm 0.02$ composition-variation cases was carried out up to timestep 327,000, as the forecasting fidelity beyond this point decreases more noticeably. It is noticeable that the framework's forecasting across all cases accelerated predictions approximately 7,200$\times$ to 62,300$\times$ relative to the phase-field simulation. Although the simulation time varies significantly depending on the configuration, the variation in the model's forecasting time is minimal and stays within the order of seconds. These results are an indication of substantial reduction in computational cost for the proposed framework across different microstructural configurations, spatial resolutions, and alloy compositions while maintaining a high level of precision.

\begin{table}[H]
\centering
\small
\caption{Phase-field vs. AE--GCN--LSTM runtime comparison}
\label{tab:runtime_comparison}
\setlength{\tabcolsep}{4pt}
\begin{tabularx}{\textwidth}{@{}lcccc@{}}
\toprule
\textbf{Configuration} &
\textbf{Timestep} &
\textbf{Simulation Time} &
\textbf{Forecast Time} &
\textbf{Speedup ($\times$)} \\
\midrule
Small FCC            & 1,644,000 & 8 h 48 min  & 1.3 s & 24,369$\times$ \\
Large FCC            & 3,000,000 & 41 h 32 min  & 2.4 s & 62,300$\times$ \\
Two-FCC              & 3,000,000 & 20 h 13 min  & 2.3 s & 31,643$\times$ \\
Five-FCC             & 3,000,000 & 16 h 7 min   & 2.2 s & 26,373$\times$ \\
$256 \times 256$     & 3,000,000 & 61 h 48 min  & 31 s  & 7,177$\times$ \\
$512 \times 512$     & 1,290,000 & 202 h 11 min & 21 s  & 34,660$\times$ \\
$\pm 0.01$ variation & 327,000   & 4 h 43 min   & 1.8 s & 9,433$\times$ \\
$\pm 0.02$ variation & 327,000   & 4 h 38 min   & 1.9 s & 8,779$\times$ \\
\bottomrule
\end{tabularx}
\end{table}

\section{Discussion}\label{sec3}

This work focused on latent graph-based AE--GCN--LSTM surrogate framework for long-horizon forecasting of microstructure evolution in multicomponent high-entropy alloys. The system studied is AlCrFeNi, containing coexisting BCC and FCC phases. The framework is an extension of our previously developed binary-alloy model into a realist quaternary, multiphase system. The proposed approach is capable of accurately forecasting the coupled evolution of four elemental concentration fields together with the phase-field order parameter. The underlying microstructural evolution is preserved during forecasting horizons extending to 3,000,000 simulation timesteps. The framework consistently predicted the evolution of previously unseen test microstructures with high predictive accuracy. 

Beyond standard forecasting, the model demonstrated strong zero-shot generalization across a broad range of previously unseen conditions without retraining, fine-tuning, or parameter adaptation. Conditions include unseen FCC precipitate sizes, spatial locations, increased numbers of precipitates, and more complex morphologies such as precipitate merging and splitting. Additionally, although the model was only trained on on $100 \times 100$ simulations containing a single FCC precipitate, it successfully generalized to more complex microstructures containing two and five FCC precipitates. Moreover, it was successfully implemented on substantially larger $256 \times 256$ and $512 \times 512$ computational domains through a patch-based strategy. The framework remained applicable to previously unseen alloy compositions as well. The approach proved transferability to unseen alloy compositions by preserving phase morphology and compositional evolution. Although larger deviations from the training composition resulted in a reduction in spatial fidelity, the framework continued to capture the underlying dynamics.

A key advantage of this method lies within its computational efficiency based on the results obtained. Across all evaluated configurations, forecasting required only a few seconds with a computational speed increase ranging from approximately $7.2 \times 10^{3}$ to $6.23 \times 10^{4}$ compared to conventional phase-field simulation. Therefore, the proposed framework is well suited for for high-throughput computational materials design and optimization.

Although the system studied focused on the high-entropy AlCrFeNi  alloy, the latent graph-based forecasting is not inherently restricted to this material system. The plan for future work is to investigate its application to additional multicomponent alloys and other materials, including lithium-ion battery systems. Furthermore, the framework discussed in this paper provides the ground for future physics-informed agentic AI capable of autonomously exploring next-generation materials discovery.

\section{Methods}\label{sec4}

\subsection{Hardware and Computational Resources}

All simulations, model training, and inference experiments were performed on a single mobile workstation (HP ZBook Power 16 G11). The system is equipped with an Intel Ultra 9-185H CPU (2.30 GHz, 16 cores), 64 GB DDR5 RAM, and 1 TB NVMe SSD storage. The workstation includes two GPU devices: NVIDIA RTX 3000 Ada Generation (8 GB dedicated VRAM) and an Intel Arc Graphics (integrated GPU). The system also contains an Intel AI Boost NPU, which was not used in the present study.

\subsection{Phase Field (Space and time dependent variables)}

Quaternary Al-Cr-Fe-Ni system, containing 2 phases FCC and BCC with the possibility of spinodal decomposition of the BCC\_B2 (ordered BCC) into BCC\_B2 +  BCC (disordered).

\subsubsection{Main (independent) variables}
\textbf{Three composition variables}: $x_{Al}(x,y,t)$,$x_{Cr}(x,y,t)$,$x_{Fe}(x,y,t)$, representing the molar fractions of Al, Cr and Fe as a function of space and time. The molar fraction of Ni, can be calculated from $x_{Ni}(x,y,t) = 1- x_{Al}(x,y,t) - x_{Cr}(x,y,t) - x_{Fe}(x,y,t)$. 
\\ \\
\textbf{Two phase-field variables} : $\phi_{BCC}(x,y,t)$ and $\phi_{FCC}(x,y,t)$, with \\
 $\phi_{BCC}=1$ and $\phi_{FCC}=0$ representing the  BCC phase, either the ordered BCC\_B2 phase or the BCC\_xx phase. The BCC\_B2 and BCC\_xx phases are distinguished based on composition.
   $\phi_{BCC}=0$ and $\phi_{FCC}=1$ representing the  FCC phase
   
\subsubsection{Dependent variables: the phase compositions}

Following the Kim-Kim-Suzuki \cite{KKS} approach, \textbf{phase compositions} are introduced
\begin{itemize}
	\item Composition BCC phase :  $x_{Al}^{BCC}(x,y,t)$,$x_{Cr}^{BCC}(x,y,t)$,$x_{Fe}^{BCC}(x,y,t)$
	\item Composition FCC phase : $x_{Al}^{FCC}(x,y,t)$,$x_{Cr}^{FCC}(x,y,t)$,$x_{Fe}^{FCC}(x,y,t)$
\end{itemize}
They are computed from the local values of the composition variables \\ ($x_{Al}(x,y,t)$,$x_{Cr}(x,y,t)$,$x_{Fe}(x,y,t)$) and phase-field variables \\ $\phi_{BCC}(x,y,t)$,  $\phi_{FCC}(x,y,t)$, assuming equal diffusion potentials in the two phases for each component and overall conservation of the local composition \\ $x_{Al}(x,y,t)$,$x_{Cr}(x,y,t)$,$x_{Fe}(x,y,t)$. 
\begin{itemize}
	\item $\tilde{\mu}_{Al}^{BCC} = \tilde{\mu}_{Al}^{FCC} = \tilde{\mu}_{Al}$,
	\item $\tilde{\mu}_{Cr}^{BCC} = \tilde{\mu}_{Cr}^{FCC} = \tilde{\mu}_{Cr}$,
	\item $\tilde{\mu}_{Fe}^{BCC} = \tilde{\mu}_{Fe}^{FCC} = \tilde{\mu}_{Fe}$
	\item $x_{Al}(x,y,t) = \phi_{BCC}(x,y,t)*x_{Al}^{BCC}(x,y,t) + \phi_{FCC}(x,y,t)*x_{Al}^{FCC}(x,y,t)$, 
	\item $x_{Cr}(x,y,t) = \phi_{BCC}(x,y,t)*x_{Cr}^{BCC}(x,y,t) + \phi_{FCC}(x,y,t)*x_{Cr}^{FCC}(x,y,t)$,
	\item $x_{Fe}(x,y,t) = \phi_{BCC}(x,y,t)*x_{Fe}^{BCC}(x,y,t) + \phi_{FCC}(x,y,t)*x_{Fe}^{FCC}(x,y,t)$.
\end{itemize}
The diffusion potentials are calculated as $\tilde{\mu}_{Al}^{BCC} = \frac{\partial G_m^{BCC}}{\partial x_{Al}^{BCC}}$,$\tilde{\mu}_{Cr}^{BCC} = \frac{\partial G_m^{BCC}}{\partial x_{Cr}^{BCC}}$, $\tilde{\mu}_{Fe}^{BCC} = \frac{\partial G_m^{BCC}}{\partial x_{Fe}^{BCC}}$,  $\tilde{\mu}_{Al}^{FCC} = \frac{\partial G_m^{FCC}}{\partial x_{Al}^{FCC}}$,$\tilde{\mu}_{Cr}^{FCC} = \frac{\partial G_m^{FCC}}{\partial x_{Cr}^{FCC}}$, $\tilde{\mu}_{Fe}^{FCC} = \frac{\partial G_m^{FCC}}{\partial x_{Fe}^{FCC}}$. 
Note that the phase compositions $x_{k}^{\rho}(x,y,t)$ ($i=Al,Cr,Fe$,$\rho=FCC,BCC$) are functions of $\phi_{BCC}$ and $\phi_{FCC}$ and the three molar fraction field $x_{Al}(x,y,t), x_{Cr}(x,y,t), x_{Fe}(x,y,t)$. 

\subsection{Thermodynamic free energy functional of the system}

\subsubsection{Overall free energy functional}

The overall free energy functional is of the form
\begin{eqnarray}
	F(\phi_{BCC},\phi_{FCC},x_{Al},x_{Cr},x_{Fe})=
	\int[f_{int}(\phi_{BCC},\phi_{FCC}) \\  + f_{chem,bulk}(\phi_{BCC},\phi_{FCC},x_{Al},x_{Cr},x_{Fe}) \\ +  \frac{\kappa}{2} [ (\nabla \phi_{BCC})^2 + (\nabla \phi_{FCC})^2] \\ + \frac{\epsilon}{2} \Phi [ (\nabla x_{Al})^2 + (\nabla x_{Cr})^2 + (\nabla x_{Fe})^2]]\ud V
	\end{eqnarray}
The matrix $\Phi$ has the same size as the system and equals 1 at grid points for which the composition is in the two-phase region BCC\_B2 + BCC, and the value 0 elsewhere. The local values of $\Phi$ are determined on every time step based on the local values of the molar fraction fields using a precalculated table listing compositions in the two-phase region obtained from Thermo-Calc.

\subsection{Chemical bulk free energy}

We define interpolation functions $h_{BCC}$ and $h_{FCC}$ as \cite{Moelans2011},
\begin{equation}
	h_{BCC}(\phi_{BCC},\phi_{FCC}) = \frac{\phi_{BCC}^2(x,y,t)}{\phi_{BCC}^2(x,y,t) + \phi_{FCC}^2(x,y,t)}
\end{equation}
and
\begin{equation}
	h_{FCC}(\phi_{BCC},\phi_{FCC}) = \frac{\phi_{FCC}^2(x,y,t)}{\phi_{BCC}^2(x,y,t) + \phi_{FCC}^2(x,y,t)}
\end{equation}
For the chemical bulk free energy, the following interpolation of the CALPHAD Gibbs energies for the BCC and FCC phases is used
\begin{eqnarray}
	f_{chem,bulk}(\phi_{BCC},\phi_{FCC},x_{Al},x_{Cr},x_{Fe}) \\ = h_{BCC}(\phi_{BCC},\phi_{FCC})\frac{G_m^{BCC}(x_{Al}^{BCC},x_{Cr}^{BCC},x_{Fe}^{BCC})}{V_m} \\ + h_{FCC}(\phi_{BCC},\phi_{FCC})\frac{G_m^{FCC}(x_{Al}^{FCC},x_{Cr}^{FCC},x_{Fe}^{FCC})}{V_m} 
\end{eqnarray}

Note that the Gibbs energies are evaluated using the phase compositions. 
The Gibbs energy functions $G_m^{BCC}(x_{Al}^{BCC},x_{Cr}^{BCC},x_{Fe}^{BCC})$ and \\ $G_m^{FCC}(x_{Al}^{FCC},x_{Cr}^{FCC},x_{Fe}^{FCC})$ and their first derivatives ($\tilde{\mu}_{Al}^{BCC}, \tilde{\mu}_{Cr}^{BCC}, \tilde{\mu}_{Fe}^{BCC} $ and $\tilde{\mu}_{Al}^{FCC}, \tilde{\mu}_{Cr}^{FCC}, \tilde{\mu}_{Fe}^{FCC} $)  are taken from the CALPHAD models for the FCC\_L12 and BCC\_B2 phase in the Thermo-Calc databases TCHEA7. Their values are calculated at different compositions $x_{Al}$, $x_{Cr}$,$x_{Fe}$, varying the molar fraction of each component as 0.01:0.01:0.99 and at a temperature T=1300 K. 

A tensor model is fitted following the method of \cite{Coutinho2020}. A Rank R=10 is assumed for the BCC phase and rank R=6 for the FCC phase. For the BCC phase, only the points outside the two-phase region $BCC + B2$ are considered in the fit. 
From the fitted tensor model, the Gibbs energies and their first and second derivatives can be computed.

The points inside the two-phase region are treated in a different way in the Cahn-Hilliard equations (see part on Cahn-Hilliard equation). The phase-fraction $f_{BCC\_B2\#1}$ is also computed with Thermo-Calc at the considered compositions. The two-phase region ($\Phi=1$ in front of the $\epsilon$ in the free energy functional) is taken at $0 < f\_{BCC_B2\#1}<1$, while the spinodal region is defined as $0.1 < f_{BCC\_B2\#1}<0.9$.     

The molar volume $V_m$ is assumed to be constant over the system and taken equal to 1e-5.

\subsection{Interface free energy}

The homogeneous part of the interfacial energy in the phase-field model has the form
\begin{eqnarray}
	f_{int}(\phi_{BCC},\phi_{FCC}) = m(\frac{\phi_{BCC}^4}{4}-\frac{\phi_{BCC}^2}{2} + \\  \frac{\phi_{FCC}^4}{4}-\frac{\phi_{FCC}^2}{2} + \frac{1}{4} + \gamma \phi_{BCC}^2 \phi_{FCC}^2)
\end{eqnarray}

Model parameters related to interface energy and thickness : $\kappa = 0.0001 e^{-8}$, $\gamma = 1.5$ and $m=1e6$.

\subsection{Evolution Equations}

\subsubsection{Evolution non-conserved phase-fields}

Allen-Cahn equation for $\phi_{FCC}$
\begin{eqnarray}
	\frac{\partial \phi_{FCC}}{\partial t} &=& -L [\frac{\partial f_{int}}{\partial \phi_{FCC}}+\frac{\partial f_{chem,bulk}}{\partial \phi_{FCC}}-\kappa \nabla^2 \phi_{FCC}] \\
    &=& -L (DFAC_{FCC})
\end{eqnarray}
with $DFAC_{FCC}$ the driving force in the Allen - Cahn equations, which is calculated in the MATLAB script as the variable \emph{driving\_froce\_AC}, and
\begin{eqnarray}
	\frac{\partial f_{int}}{\partial \phi_{FCC}} = 	m[\phi_{FCC}^3 - \phi_{FCC} + 2\gamma \phi_{BCC}^2\phi_{FCC} ]
\end{eqnarray}
\begin{eqnarray}
	\frac{\partial f_{chem,bulk}}{\partial \phi_{FCC}} = 
	\frac{\partial h_{BCC}}{\partial \phi_{FCC}} \frac{G_m^{BCC}}{V_m} 
	+ \frac{\partial h_{FCC}}{\partial \phi_{FCC}} \frac{G_m^{FCC}}{V_m} \\
	+ h_{BCC}\frac{1}{V_m}[\frac{\partial G_m^{BCC}}{\partial x_{Al}^{BCC}}\frac{\partial x_{Al}^{BCC}}{\partial \phi_{FCC}} +
	\frac{\partial G_m^{BCC}}{\partial x_{Cr}^{BCC}}\frac{\partial x_{Cr}^{BCC}}{\partial \phi_{FCC}} + 
	\frac{\partial G_m^{BCC}}{\partial x_{Fe}^{BCC}}\frac{\partial x_{Fe}^{BCC}}{\partial \phi_{FCC}}] \\
	+ h_{FCC}\frac{1}{V_m}[\frac{\partial G_m^{FCC}}{\partial x_{Al}^{FCC}}\frac{\partial x_{Al}^{FCC}}{\partial \phi_{FCC}} +
	\frac{\partial G_m^{FCC}}{\partial x_{Cr}^{FCC}}\frac{\partial x_{Cr}^{FCC}}{\partial \phi_{FCC}} + 
	\frac{\partial G_m^{FCC}}{\partial x_{Fe}^{FCC}}\frac{\partial x_{Fe}^{FCC}}{\partial \phi_{FCC}}]
\end{eqnarray}
Allen-Cahn equation for $\phi_{BCC}$
\begin{eqnarray}
	\frac{\partial \phi_{BCC}}{\partial t} &=& -L [\frac{\partial f_{int}}{\partial \phi_{BCC}}+\frac{\partial f_{int}}{\partial \phi_{BCC}}-\kappa \nabla^2 \phi_{BCC}]\\
    &=& -L (DFAC_{BCC})
\end{eqnarray}
with $DFAC_{BCC}$ the driving force in the Allen - Cahn equations, which is calculated in the matlab script as the variable \emph{driving\_froce\_AC}, and
\begin{eqnarray}
	\frac{\partial f_{int}}{\partial \phi_{BCC}} = 	m[\phi_{BCC}^3 - \phi_{BCC} + 2\gamma \phi_{BCC}\phi_{FCC}^2 ]
\end{eqnarray}
\begin{eqnarray}
	\frac{\partial f_{chem_bulk}}{\partial \phi_{BCC}} = 
	\frac{\partial h_{BCC}}{\partial \phi_{BCC}} \frac{G_m^{BCC}}{V_m} 
	+ \frac{\partial h_{FCC}}{\partial \phi_{BCC}} \frac{G_m^{FCC}}{V_m} \\
	+ h_{BCC}\frac{1}{V_m}[\frac{\partial G_m^{BCC}}{\partial x_{Al}^{BCC}}\frac{\partial x_{Al}^{BCC}}{\partial \phi_{BCC}} +
	\frac{\partial G_m^{BCC}}{\partial x_{Cr}^{BCC}}\frac{\partial x_{Cr}^{BCC}}{\partial \phi_{BCC}} + 
	\frac{\partial G_m^{BCC}}{\partial x_{Fe}^{BCC}}\frac{\partial x_{Fe}^{BCC}}{\partial \phi_{BCC}}] \\
	+ h_{FCC}\frac{1}{V_m}[\frac{\partial G_m^{FCC}}{\partial x_{Al}^{FCC}}\frac{\partial x_{Al}^{FCC}}{\partial \phi_{BCC}} +
	\frac{\partial G_m^{FCC}}{\partial x_{Cr}^{FCC}}\frac{\partial x_{Cr}^{FCC}}{\partial \phi_{BCC}} + 
	\frac{\partial G_m^{FCC}}{\partial x_{Fe}^{FCC}}\frac{\partial x_{Fe}^{FCC}}{\partial \phi_{BCC}}]
\end{eqnarray}

Parameter values : $L = 1.1922$e-4

\subsubsection{Evolution of conserved molar fraction fields : diffusion and spinodal decomposition of BCC}

Cahn-Hilliard evolution equations of the molar fraction fields $x_k(x,y,t)$

\begin{eqnarray}
	\frac{\partial x_k}{\partial t} = V_m \nabla \cdot [ M_{bulk}  \nabla [\tilde{\mu}_k^* - 2V_m\Phi
	\epsilon \nabla^2 x_k]]
\end{eqnarray}
with, in the spinodal region
\begin{equation}
	\tilde{\mu}_k^* = (\tilde{\mu}_k)_{CALPHAD} -W_{spin}x_k
\end{equation}
to mimic the uphill driving force for diffusion in the spinodal region, and in the 2-phase region outside the spinodal region
\begin{equation}
	\tilde{\mu}_k^* = (\tilde{\mu}_k)_{CALPHAD} +W_{spin}x_k
\end{equation}
to drive the diffusion towards the outside of the 2-phase region. For al compositions, 
\begin{equation}
	\tilde{\mu}_l = \frac{\partial G_m^{BCC}}{\partial x_l^{BCC}} = \frac{\partial G_m^{FCC}}{\partial x_l^{FCC}}
\end{equation}
as prescribed by the Kim-Kim-Suzuki equations.

Parameter values : $M_{bulk}=1e-19$, $W_{spin}=0.001*\frac{\partial^2 G_m^{BCC}}{\partial (x_k^{BCC})^2}$, $\epsilon = 0.1*\frac{\partial^2 G_m^{BCC}}{\partial (x_k^{BCC})^2}$.

In the Matlab script, the variables $\tilde{\mu}_k^* - 2V_m\Phi \epsilon \nabla^2x_k$ are computed and stored in the variable \emph{diff\_pot}. Moreover, the variables \emph{flux\_1} and \emph{flux\_1} contain the opposite of the fluxes of the different components in directions 1 and 2. These are calculated as 
\begin{eqnarray}
-J_1 = V_m   M_{bulk}  \frac{\ud}{\ud x_1} [\tilde{\mu}_k^* - 2V_m\Phi
	\epsilon \nabla^2 x_k]
\end{eqnarray}
and
\begin{eqnarray}
-J_2 = V_m   M_{bulk}  \frac{\ud}{\ud x_2} [\tilde{\mu}_k^* - 2V_m\Phi
	\epsilon \nabla^2 x_k]
\end{eqnarray}

\subsection{Numerical and system parameters}

$\Delta x = 5$e-8 m. 
The time step is taken adaptive. The time at a certain simulation step is given in the variable \emph{time}. 

Periodic boundary conditions. The diffuse interface width is 3.1e-7 m, which means approximately 6 to 7 grid points over a diffuse interface. 

\subsection{Deep Learning Framework}

\subsubsection{Data Preparation}

The KKS phase-field simulation collected data comprise microstructure snapshots that include the phase-field order parameter $\eta$ and the elemental concentration fields for the AlCrFeNi alloy. The dataset was generated with the nominal composition of Al = 0.17, Cr = 0.33, Fe = 0.17, and Ni = 0.33. The phase-field simulation was done until 3,000,000 timesteps and snapshots were saved every 3,000 timesteps. A total of 1,000 microstructure snapshots were generated, each with a spatial resolution of 100 × 100 pixels. A (80/10/10) split was performed on the The dataset, dividing it into 800 training, 100 validation, and 101 testing samples.

The developed deep learning framework utilizes the same time adaptive variable that is exported from the phase-field simulation. The retained temporal intervals preserve the physical time scale of microstructure evolution. To ensure that every sample had the same input format, the four elemental compositions were arranged into a single standardized composition representation.

Subsequently, the concentration and phase-field parameters were combined into a five-channel microstructure representation. This five-channel representation consists of of Al, Cr, Fe, Ni, and $\eta$. Before training, the concentration fields were normalized to prevent information leakage. The phase-field variable was retained in its original physical range. The microstructure snapshots were organized in chronological order and divided into training, validation, and testing subsets.

For model development, consecutive microstructure states were grouped into short temporal sequences of length 3. The grouped temporal sequences serve as an input context for learning the evolution dynamics. This preprocessing procedure produces a structured spatiotemporal dataset that is suitable for latent-space training and forecasting.

\subsubsection{Multi-Head Latent Representation Framework}

To facilitate learning the evolution of microstructures in a compact format, a multi-head convolutional autoencoder \cite{hinton} was developed. The model receives input as a five-channel state consisting of the four elemental concentration fields (Al, Cr, Fe, Ni) and the phase-field order parameter $\eta$. The encoder first compresses each $100\times100\times5$ microstructure into a $80\times80\times128$ latent representation using convolutional layers followed by adaptive average pooling \cite{bank}. This process is performed while preserving the dominant spatial features of the composition and phase morphology.

A key design consideration is the physical distinction between concentration fields and phase-field variables. This is fundamentally important, as they represent different physical information. The concentration fields describe the local chemical composition and must satisfy conservation constraints. The phase-field variable represents the microstructural morphology and follows a different physical evolution mechanism. To ensure that this distinction is preserved, the decoder reconstructs the quantities through separate output branches. One branch predicts the four elemental concentration fields, while a second independent branch predicts the phase-field order parameter. Through this process, leakage and mixing of the distinct variables are prevented since the model does not treat them as interchangeable quantities. In order to preserve the average elemental composition, conservation constraints are enforced exclusively on the concentration branch. However, the phase-field branch is optimized independently using a binary classification that is used for phase optimization.

During training, each batch consists of 4 sequences of length 3, which are reshaped into 12 individual microstructures for autoencoder training. The model is optimized using AdamW with a learning rate of $2\times10^{-4}$ and weight decay of $10^{-5}$. Training is performed for up to 100 epochs with early stopping using a patience of 70 epochs. For model training, a composite loss function is implemented consisting of concentration reconstruction error, structural similarity preservation, phase-field reconstruction loss, and elemental conservation penalties. To help with the accurate reconstruction of the concentration field, the Mean squared error (MSE) and structural similarity index (SSIM) \cite{wang} were applied, and for the phase-field predictions a binary cross-entropy loss was used. additionally, the conservation term was evaluated to maintain consistency with the underlying thermodynamic system. The objectives implemented promote an accurate reconstruction and prediction of compositional and morphological features.

\begin{figure}[H]
    \centering
    \includegraphics[width=1.0\linewidth]{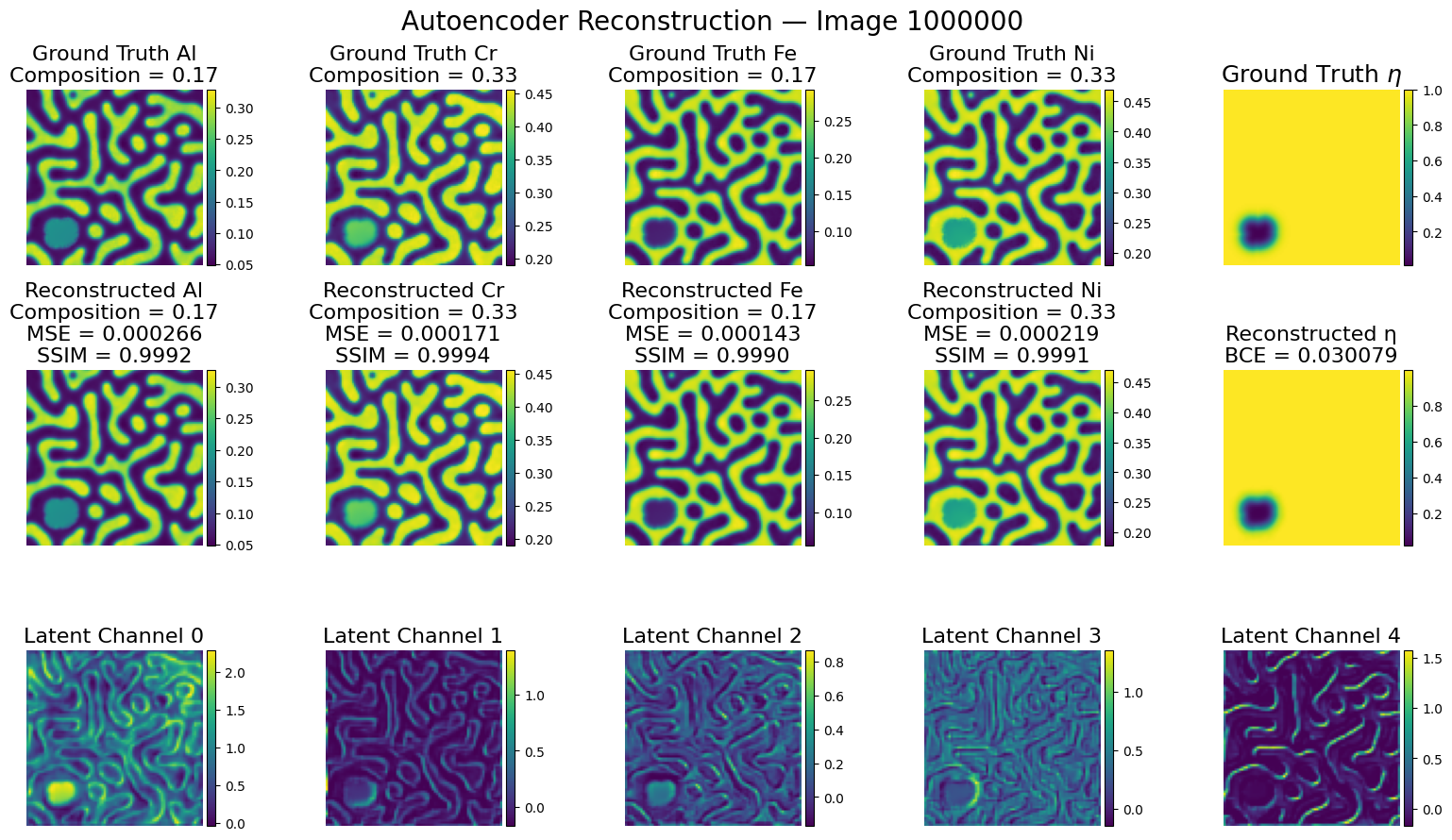}
    \caption{Encoding of input microstructures into a latent representation. Reconstruction quality (SSIM and MSE), alloy compositions, and representative latent channels are shown.}
    \label{fig:autoencoder_visualization}
\end{figure}

Upon training the autoencoder, all microstructure sequences are projected into the latent space by the encoder network. This transformation enables a compact representation of the coupled concentration and phase fields while retaining important spatiotemporal information for forecasting. This encoding process is applied separately to the training, validation, and test datasets. The obtained latent representations along with the corresponding microstructures from the datasets are used to train the downstream GCN--LSTM forecasting. As shown in Fig.~\ref{fig:autoencoder_visualization} shows the encoding of the input microstructures into a latent representation prior to graph construction for downstream GCN--LSTM forecasting. In the image, the reconstruction quality with SSIM, and MSE is included along with the predicted and ground-truth alloy compositions and latent channels learned by the autoencoder.

\subsubsection{Graph Construction from Latent Microstructure Fields}

Following the latent transformation, The encoded fields are converted into graph-structured representations. In the feature map, each spatial location is treated as a graph node connected to the neighboring nodes. The pattern follows a four-neighbor grid connectivity corresponding to the up, down, left, and right directions. By implementing this construction, the local spatial relationships of the microstructure are preserved while allowing graph-based message passing over the latent domain \cite{thomas}. Since the encoder produces a latent representation of $80 \times 80 \times 128$, it is therefore converted into a graph containing $6,\!400$ nodes for downstream GCN processing.

\begin{figure}[H]
    \centering
    \includegraphics[width=1.0\linewidth]{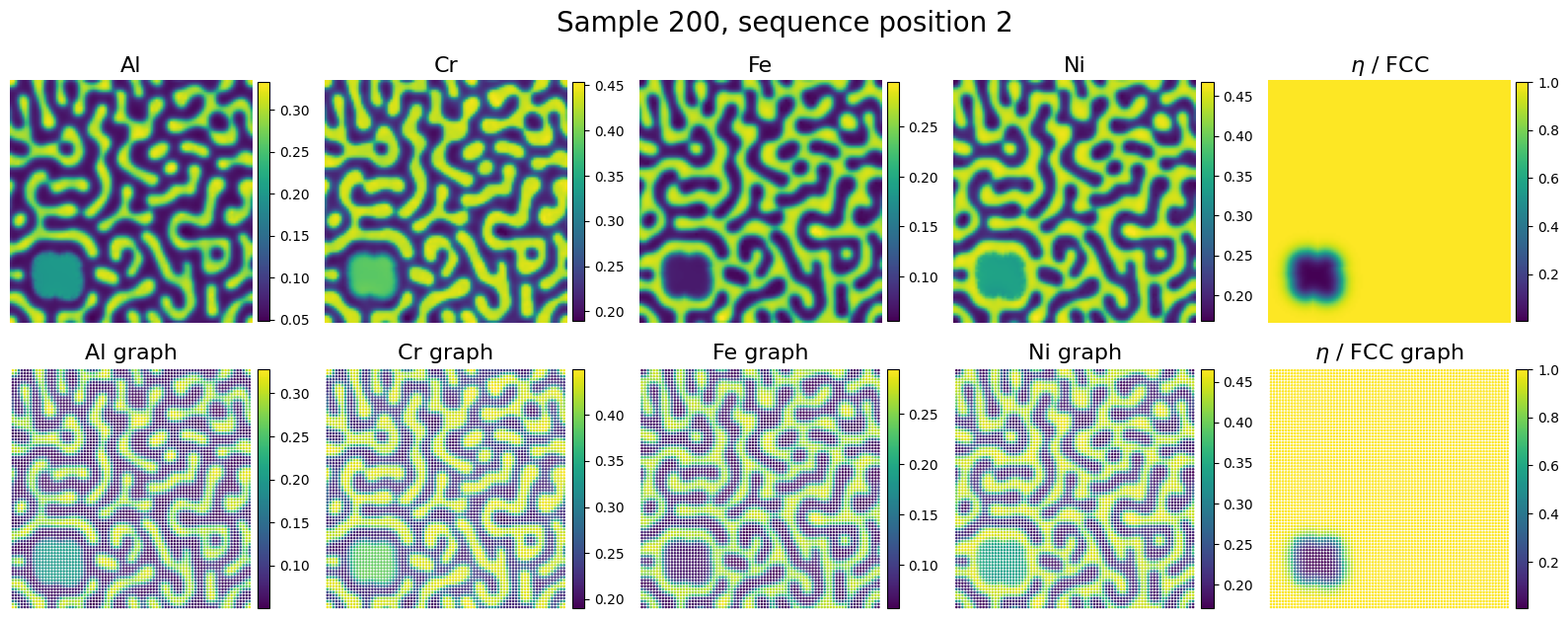}
    \caption{Graph representation of the encoded latent microstructure. Nodes represent latent spatial locations, and edges connect neighboring nodes.}
    \label{fig:graph_construction}
\end{figure}

To retain the temporal order of the sequences, a separate graph is constructed at every time step. For each graph representation, node features are defined by the local latent vector at each spatial position. Therefore, the resulting sequence of graphs provides a compact spatiotemporal representation of the evolving microstructure. Ground-truth microstructure snapshots are also paired with these graph sequences to support supervised training of the downstream GCN--LSTM forecasting model. As illustrated in Fig.~\ref{fig:graph_construction}, the encoded latent representation is converted into a graph. Each node corresponds to a spatial location in the latent feature map, and edges connect neighboring nodes to preserve local spatial relationships.

\subsubsection{Graph-Based Temporal Forecasting}

The model combines graph convolutional networks (GCNs) and long short-term memory (LSTM) \cite{lstm} networks to learn the spatiotemporal evolution of latent microstructure representations. Upon conversion of the latent feature map into a graph, at each time step the graph is processed by a stack of graph convolutional layers. 

As stated earlier, each latent graph contains $6,\!400$ nodes, while each node is represented by a $128$-dimensional feature vector. Consequently, the graph is processed by four graph convolutional layers. The first layer projects the $128$ input features into a $256$-dimensional hidden representation. The remaining three layers maintain the same hidden dimension. Next step is followed by Batch normalization \cite{batch}, ReLU activation, and dropout between graph convolutional layers. This process helps to improve training stability and reduce overfitting. With this setup, each node embedding utilizes information from its surrounding to enable the model to learn local microstructural dynamics in latent space.

Subsequently, node embeddings from all time steps are organized into temporal sequences and passed to recurrent networks to capture the temporal evolution of microstructures. The LSTM architecture is composed of two distinct branches, each with a hidden dimension of 256. The first branch is implemented for the compositional evolution and predicts the four elemental concentration fields. The second branch models phase evolution and predicts the order parameter $\eta$. Prior to prediction, Layer Normalization \cite{layer} and Dropout \cite{dropout} are applied to the outputs of both recurrent branches.

An important feature of this architecture is the separation between concentration and phase-field forecasting. Although the two branches operate on the same latent representation, they maintain independent temporal dynamics and prediction heads. This design prevents the entanglement of conserved compositional quantities from the non-conserved phase-field variable. To further reduce the possibility of leaking features between the composition and phase-field variables, an orthogonality constraint is imposed. This orthogonality encourages each branch to learn complementary physical information.

The concentration branch outputs the four elemental concentration fields, whereas the phase branch outputs phase logits. The phase logits are then converted into phase probabilities. Conservation constraints are only applied to the concentration head by enforcing an agreement between the predicted and the ground truth mean compositions. This method ensures that the elemental concentrations are retained and consistent throughout the process.

Training is performed using a composite function consisting of concentration reconstruction loss, structural similarity preservation, phase-field prediction loss, and conservation penalties. This architecture integrates graph-based spatial learning with recurrent temporal modeling along with output separation between the main variables. These objectives are combined unified framework for long-horizon microstructure forecasting. The model was trained using the AdamW optimizer with a learning rate of $2\times10^{-4}$, and was performed for up to $100$ epochs with an early-stopping patience of $50$ epochs.

\section*{Acknowledgment}

This work was supported by the European Research Council (ERC) under the EU Horizon 2020 program (Grant No. 101123107, mTWIN – Innovative digital twin concept of complex microstructure evolution in multi-component materials) and by the Research Foundation – Flanders (FWO, Grant No. K805224N, “Exploring pathways to bring microstructure simulations closer to industrial needs.

\end{document}